\documentclass[conference]{IEEEtran}
\IEEEoverridecommandlockouts
\usepackage{cite}
\usepackage{amsmath,amssymb,amsfonts}
\usepackage{algorithmic}
\usepackage{graphicx}
\usepackage{lipsum}
\usepackage{textcomp}
\usepackage{xcolor}
\usepackage{comment}
\usepackage{tabularx}
\usepackage{setspace}
\usepackage{multirow}  
\usepackage{array}
\usepackage{float}
\usepackage{comment}
\usepackage[hidelinks]{hyperref}
\usepackage{enumitem}
\usepackage{amsmath}
\usepackage{siunitx}
\usepackage[utf8x]{inputenc} 
\usepackage{float}  
\usepackage{pgfplots}
\pgfplotsset{compat=1.18}
\usepackage{upgreek}
\usepackage{arydshln}

\def\BibTeX{{\rm B\kern-.05em{\sc i\kern-.025em b}\kern-.08em
    T\kern-.1667em\lower.7ex\hbox{E}\kern-.125emX}}
\begin{document}

\title{Low Complexity Learning-based \\ Lossless Event-based Compression\\

\thanks{This work was supported by the RayShaper SA, Valais, Switzerland, through the Project entitled Event Aware Sensor Compression, and by the Fundação para a Ciência e a Tecnologia (FCT), Portugal, through the Project entitled Deep Compression: Emerging Paradigm for Image Coding under Grant PTDC/EEI-COM/7775/2020.}
}

\author{
   \IEEEauthorblockN{Ahmadreza Sezavar\IEEEauthorrefmark{1}, Catarina Brites\IEEEauthorrefmark{2}, Jo\~{a}o Ascenso\IEEEauthorrefmark{3}}
   \IEEEauthorblockA{\IEEEauthorrefmark{1}\IEEEauthorrefmark{3}Instituto Superior Técnico,\IEEEauthorrefmark{2}Instituto Universitário de Lisboa (ISCTE-IUL), \IEEEauthorrefmark{1}\IEEEauthorrefmark{2}\IEEEauthorrefmark{3}Instituto de Telecomunicações}
    \IEEEauthorblockA{\IEEEauthorrefmark{1}ahmadreza.sezavar@lx.it.pt\IEEEauthorrefmark{2}catarina.brites@lx.it.pt\IEEEauthorrefmark{3}joao.ascenso@lx.it.pt}
}

\maketitle

% ===========================================================
% abstract
% ===========================================================
\begin{abstract}
Event cameras are a cutting-edge type of visual sensors that capture data by detecting brightness changes at the pixel level asynchronously. These cameras offer numerous benefits over conventional cameras, including high temporal resolution, wide dynamic range, low latency, and lower power consumption. However, the substantial data rates they produce require efficient compression techniques, while also fulfilling other typical application requirements, such as the ability to respond to visual changes in real-time or near real-time. Additionally, many event-based applications demand high accuracy, making lossless coding desirable, as it retains the full detail of the sensor data. Learning-based methods show great potential due to their ability to model the unique characteristics of event data thus allowing to achieve high compression rates. This paper proposes a low-complexity lossless coding solution based on the quadtree representation that outperforms traditional compression algorithms in efficiency and speed, ensuring low computational complexity and minimal delay for real-time applications. Experimental results show that the proposed method delivers better compression ratios, i.e., with fewer bits per event, and lower computational complexity compared to current lossless data compression methods.
\end{abstract}

\begin{IEEEkeywords}
event cameras, compression, lossless, quadtree, rice coding, probability model prediction
\end{IEEEkeywords}

% ===========================================================
% Introduction
% ===========================================================

\section{Introduction}
\begin{comment}
\end{comment}

\par An event camera, also known as a dynamic vision sensor (DVS), is a type of sensor that differs from traditional cameras in the way it captures visual information. Instead of capturing entire frames at fixed intervals like a conventional camera, an event camera operates at the pixel level by detecting changes in brightness (events). This results in a highly efficient method of capturing visual data with extremely low latency and high dynamic range, making it ideal for applications requiring fast responses and minimal computational resources. Event cameras are important because they offer significant advantages in tasks such as high-speed motion tracking, low-latency robotics, and dynamic scene analysis, where traditional cameras may struggle, especially in high-speed motion scenes and scenes with uncontrolled illumination conditions.

\par
\begin{figure}[h]
\includegraphics[scale=0.6]{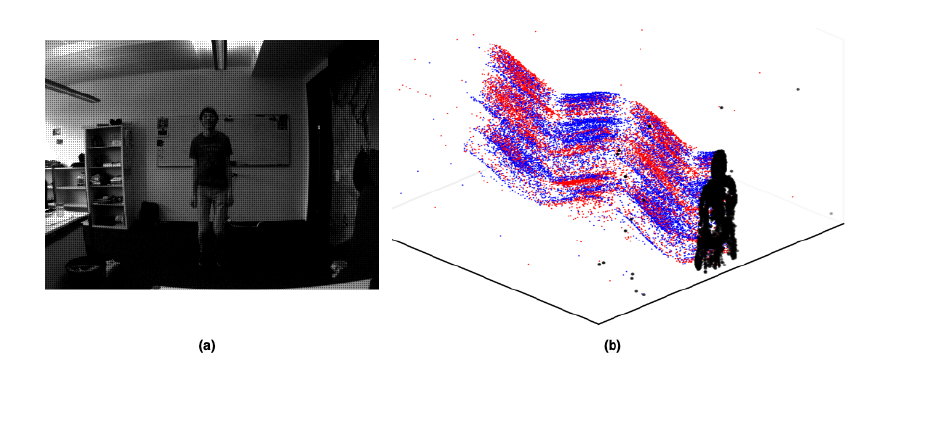}
\centering
\caption{Visualization of event data in 3D space: (a) Ground truth image captured by the camera (b) A set of events produced by an event camera, events in red (positive polarity) and blue (negative polarity) with the first events in black to show scene structure.}
\label{fig:event}
\end{figure}

 An event is a single occurrence captured by a pixel within the event camera's sensor, represented as a 4D tuple $(t_s, x, y, p)$, where $t_s$ denotes the precise timestamp of the event's occurrence, $(x, y)$ specifies the spatial coordinates within the sensor's pixel array, and $p$ indicates the polarity of the brightness change (increase or decrease) at that location. Figure \ref{fig:event} depicts an event sequence in 3D space $(x, y, t_s)$ and illustrates several benefits of event-based data, such as time-space continuity, absence of blur, and rapid acquisition to minor changes in the scene.

\par Lossless compression of an event sequence is highly desirable to preserve the integrity of the captured data. Event cameras generate a continuous stream of asynchronous events, each representing a precise spatio-temporal change in brightness. Losing any events or reducing its accuracy could lead to missed details or inaccuracies in the processing of visual scenes, potentially compromising the performance of computer vision algorithms and other applications relying on this type of data.

\par Furthermore, many applications that leverage event data, such as autonomous vehicles, robotics, and augmented reality systems, demand real-time processing and analysis of visual information. Consequently, fast and low-latency encoding and decoding of the event data are crucial. Any significant delays in encoding or decoding the event sequence could result in delayed decision-making or outdated information, potentially leading to safety hazards or sub-optimal performance. In the context of autonomous vehicles, for instance, real-time processing of event data is essential for timely detection and tracking of obstacles, pedestrians, and other moving objects on the road. Even a brief delay in processing this information could have severe consequences, as vehicles traveling at high speeds cover significant distances in a short time. Real-time encoding and decoding of event data ensures that the vehicle's perception and decision-making systems have access to the most up-to-date visual information, enabling prompt and appropriate responses to dynamic road conditions.

\par In this context, this work aims to introduce a novel lossless event data compression method that leverages deep learning techniques (which are also referred as learning-based), while maintaining low computational requirements and supporting low-latency streaming. The focus is on scenarios where real-time execution of an application from a live sensor stream takes place. In this case, event data is instantly encoded by the sensor, streamed to the application processor, and then decoded and processed in real-time by the application. The proposed method organizes the event sequence into a 2D event frame, which stores events that occur at spatial coordinates $(x, y)$ at a given time instance $t_s$. To improve data representation efficiency, a quadtree structure is used for the first time in this work to provide adaptive partitioning of the 2D frame of events. This structure enables a more compact binary representation of event data, which is then processed using simple but yet efficient entropy coding techniques. A key contribution of this work is the development of a deep neural network that learns to model the data distribution, thereby optimizing the performance of the entropy coder. Experimental results show that this method overcomes traditional lossless data compression techniques, achieving improvements in both computational complexity (especially at the encoder) and coding efficiency.

\par The rest of the paper is structured as follows: Section \ref{sec:related} briefly reviews background work on lossless event data compression, while the proposed compression method is described in Section \ref{sec:proposed}. Performance evaluation is presented and analysed in Section \ref{sec:performance}, and finally, Section \ref{sec:conclusions} concludes the paper.

% ===========================================================
% Related works
% ===========================================================
\section{Related Work}\label{sec:related}

\par In recent years, several studies have been published in the area of lossless compression for event data. Bi et.al. \cite{bi2018spike} introduced a pioneering method, proposing a cube-based coding framework that organizes event sequences into a 3D space-time volume followed by adaptive partitioning strategies to leverage spatio-temporal correlations. The event location, timestamp, and polarity data are independently encoded with two prediction modes, address-prior mode and time-prior mode, to accommodate different event spatial distributions. Schiopu and Bilcu \cite{schiopu2022lossless} presented a compression method for encoding 2D event frames (temporal aggregation of event data) by representing them as Event Map Images (EMI) and polarity vectors. Similarly, Khan et al. \cite{khan2020time} proposed a coding strategy that combines time aggregation and HEVC video coding for lossless compression of polarity-based event frames. 

\par Recent advancements in lossless coding methods for event data have leveraged existing point cloud compression techniques. Martini et al. \cite{martini2022lossless} and Huang et al. \cite{huang2023event} independently proposed similar approaches that represent event sequences as 3D points in a space-time volume. In both methods, each point's coordinates correspond to pixel location $(x, y)$ and timestamp $t_s$, with polarity as an attribute. Martini et al. \cite{martini2022lossless} splits the event sequence into two distinct 3D point clouds, according to the polarity, i.e. a positive event point cloud and a negative event point cloud, which are compressed separately. Huang et al. \cite{huang2023event} first splits the event sequence into multiple segments with a fixed number of events each, and then creates multiple sets of two 3D point clouds, one 3D point cloud for each polarity. Huang et al. \cite{huang2023event} also studied the case where the polarity is regarded as an extra attribute. Both methods apply Geometry-based Point Cloud Compression (G-PCC) to encode the geometric information $(x, y, t_s)$ of each polarity-based point cloud; Huang et al. \cite{huang2023event} applies a transformation to the points' coordinates before G-PCC coding to have a more appropriate range of the coordinates' values. These approaches demonstrate the potential of adapting standardized point cloud compression solutions for efficient event data encoding.

In \cite{gong2023lossless} a lossless compression approach was proposed that exploits both spatial and temporal redundancies in the event data. This method employs a multi-level dictionary structure and operates in three key stages: i) spatial domain partitioning to reduce coordinate data size, ii) compact representation of coordinates with a multi-level dictionary structure, and iii) temporal compression via differential encoding of timestamps. The algorithm was evaluated on real and synthetic datasets and outperformed other traditional and event-based compression methods like Huffman, Zlib, and EVT3.0. In \cite{schiopu2022low}, the so-called low-complexity lossless compression of asynchronous event sequence (LLC-ARES) method was introduced, which is suitable for hardware implementation in event signal processing (ESP) chips. This method involves rearranging the input event sequence using the Same-Timestamp (ST) representation and applying the triple threshold partition (TTP) algorithm. Residual errors are computed by dividing the input range into several coding ranges arranged at concentric distances from an initial prediction. 

In \cite{schiopu2023entropy}, the LLC-ARES solution was extended by adapting the TTP algorithm to use entropy coding-based techniques. Different from the previous solution, a new x-coordinate prediction strategy is used. The entropy coding-based technique employed use a Laplacian estimator to model probabilities and a list of adaptive Markov models, demonstrating improved runtime performance compared to LZMA. This method, which involves aggregating events and processing rearranged event sequences, is more suitable for applications that do not require real-time asynchronous event processing.

% ===========================================================
% Methodology
% ===========================================================
\section{Low-Complexity Learning-based Lossless Event data Compression}\label{sec:proposed}

\par The proposed Low-Complexity Learning-based Lossless Event data Compression framework, hereinafter referred to as the LC-LLEC framework, is based on three key operations: i) adaptive partitioning of the 2D spatial coordinates of the input event data using a quadtree data structure; ii) a simple custom-designed neural network to generate the probability mass function (PMF) that characterizes the statistics of the source data for the current frame; and iii) Rice coding as the entropy coding engine. This solution was developed for real-time streaming of event data, where strict delay and processing constraints must be met.

\par
\begin{figure}[h]
\includegraphics[scale=0.6]{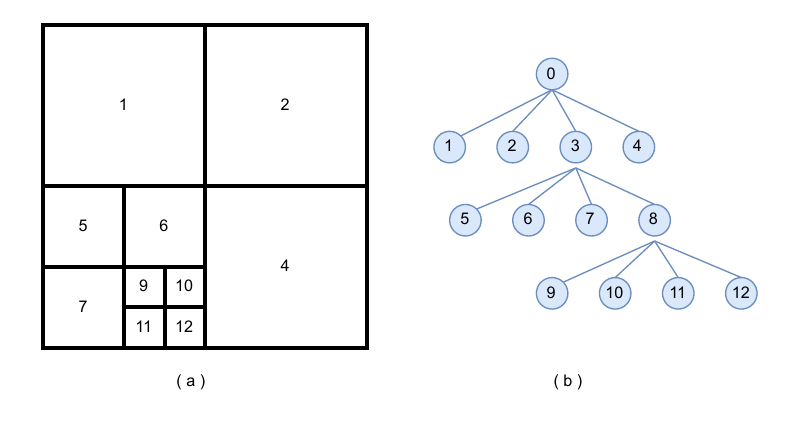}
\centering
\caption{Quadtree representation: (a) quadrants, and (b) nodes.}
\label{fig:quadtree}
\end{figure}

 As shown in Figure \ref{fig:quadtree}, a quadtree is particularly suitable for sparse data representation due to its hierarchical structure, which efficiently partitions the 2D space into quadrants, allowing for effective management of spatial data. In a quadtree, each node represents a squared region of the 2D space, which is recursively subdivided into four equal quadrants until a specified resolution is reached or until a node contains a fixed number of data points. This approach enables the quadtree to adaptively allocate resources only to areas of interest, making it highly efficient for sparse event data sequences where most of the space may be unoccupied. Occupancy in the context of a quadtree refers to the presence of data points within a given region of the tree. To extract occupancy information from the tree, the quadtree is traversed and the number of data points within each quadrant node is computed; another possibility is to just check whether a node is empty or filled. The occupancy symbols of the quadrants can be represented as a sequence of nibbles (four binary digits), where the digit ‘1’ indicates a non-empty quadrant and the digit ‘0’ signifies an empty quadrant. This binary representation of the quadtree structure may be either lossy or lossless, depending on the maximum depth of the quadtree, which determines the maximum number of recursive subdivisions that can be performed.

\par Rice coding is an efficient data compression technique particularly suitable for encoding integers. This method is especially beneficial for sparse data, as it effectively compresses small values (around zero), providing efficient compression when these values are frequent, thus, this entropy coding method is especially useful when the data is skewed towards smaller values. In this case, due to the nature of event data, the occupancy nibbles do not follow any specific distribution and thus Rice coding was combined with frequency substitution, a technique used in data compression where the occupancy nibbles (input symbols) are mapped to codes based on their frequency of occurrence. The idea is to replace more frequent symbols with shorter codes and less frequent symbols with longer codes. In this work, a simple neural network predicts the PMF to be used, i.e. the probability of occurrence for each occupancy symbol, from which the substitution table is obtained.

\subsection{Architecture and Walkthrough}

\begin{figure*}[t!]
\centering
\includegraphics[scale=1.0]{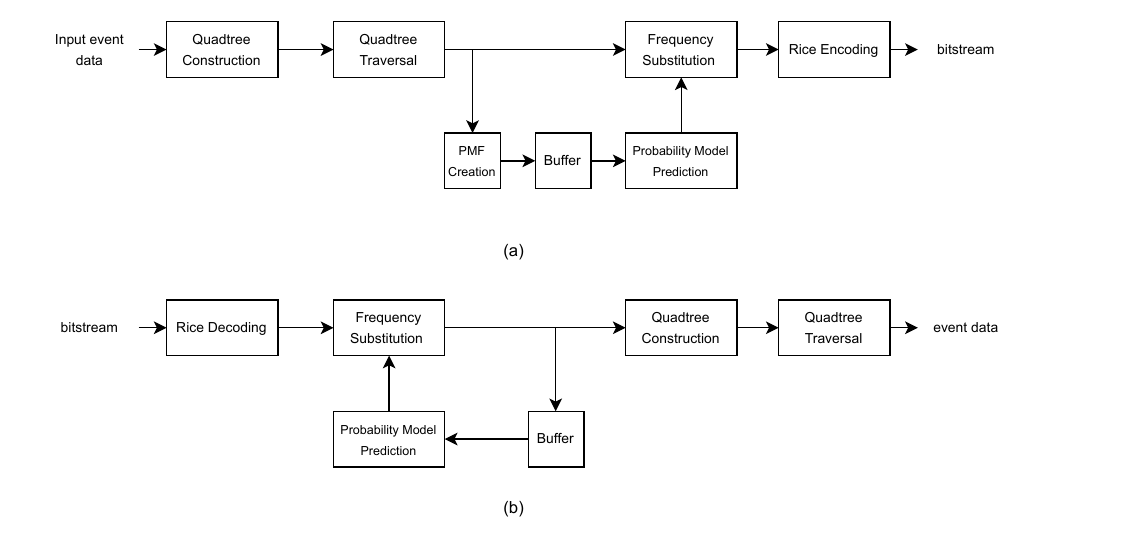}
\caption{LC-LLEC framework: (a) encoder, and (b) decoder architectures.}
\label{fig:Main_arch}
\end{figure*}

Figure \ref{fig:Main_arch} illustrates the proposed LC-LLEC architecture for both the encoder and decoder. Before applying the proposed LC-LLEC solution, the input event sequence is pre-processed. At each timestamp, the events are sorted based on their x-coordinates, and then by their y-coordinates. These sorted events form a single coding unit, like a frame in video coding scenarios, which means that the coding delay is not higher than the time precision of the timestamp, usually \SI{1}{\micro\second}. Moreover, the full timestamp values do not need to be transmitted to the decoder since the occupancy of all coding units are sent sequentially, which just corresponds to a small header within each coding unit. A single vector is also extracted, containing only the polarity of each event (+1 for positive events, -1 for negative events) for each coding unit. Not that all events on a coding unit have the same timestamp. After binarization of this polarity vector, it is transmitted directly to the decoder (without entropy coding). The $(x, y)$ coordinates will be transformed and compressed independently of the polarity vector.

The main modules of the LC-LLEC encoder and decoder architectures are described next. The set of steps presented below are performed for every coding unit (or frame), which represents the event data associated with a given timestamp of the sequence.
\begin{enumerate}[leftmargin=*]
    \item \textbf{Quadtree Construction}: A quadtree is created from the input pre-processed data or from the decoded occupancy nibbles (decoder side). The depth of the quadtree is determined by $D=log_2(\max(x_{\text{max}},y_{\text{max}}))$, where $x_{\text{max}}$, and $y_{\text{max}}$ are the maximum value that can be assumed by spatial coordinates $x$ and $y$, respectively. The computed depth $D$ guarantees that this process is lossless, i.e. each leaf node of the quadtree characterizes a spatial location (and thus an event) without any ambiguity.
    \item \textbf{Quadtree Traversal}: At the encoder, the quadtree is traversed, and the occupancy nibbles at each level are computed. At the decoder, the quadtree is also traversed until all the leaf nodes are reached, pruning whenever is appropriate. This allows to obtain the decoded event data, which is mathematically identical to the input data. 
    \item \textbf{PMF Creation and Buffer}: From the occupancy nibbles, the probability mass function (PMF) is created. This process is performed for all coding units; a buffer stores the $N$ past PMFs, which are associated with the $N$ past coding units. Note that past spatial coordinates of the events generated are not actually stored, just the PMFs, which represent the probability of occurrence for every occupancy nibble (16 values in total for every timestamp).
    \item \textbf{Probability Model Prediction}: In this step, the probability mass function for the current coding unit is predicted using a neural network based on statistics (PMFs) of past coding units. The past 10 PMFs calculated in the previous step are used as input to predict the PMF of the coding unit to be compressed, which needs to be known at both encoder and decoder. The  proposed neural network architecture is presented in detail in Section \ref{sec:net_arch}.
    \item \textbf{Frequency Substitution}: Since in Rice coding smaller integers are assigned to shorter codewords, the objective of this step is to ensure that the most frequent symbols receive the shortest codewords, thereby enhancing overall coding efficiency. Encoding the occupancy stream using frequency substitution is a two-pass process. The first pass is to create a lookup table of substitution codes according to the PMF obtained in the previous step. Then a second pass is made, where an encoded data stream is generated using the lookup substitution table. The first pass is straightforward, it corresponds to the creation of a lookup table with every possible symbol, sorted according to their PMF value (from highest to lowest). The second pass corresponds to the substitution of a symbol for its position in the frequency table. The most frequent occupancy symbols, positioned at the top of the frequency table, receive lower position values and consequently shorter Rice codes. The PMF values from the previous step determine this frequency-based ordering of occupancy symbols. 
    \item \textbf{Rice Coding}: In the final stage, Rice coding is applied to each occupancy nibble, producing the bitstream. Rice coding is a compression method that effectively encodes small non-negative integers using straightforward bit operations, attributing smaller codewords for lower-valued symbols. This technique is well-suited for the task at hand because it offers a balance between compression efficiency and computational simplicity. This step is explained in more detail in Section \ref{sec:rice}.
\end{enumerate}

\subsection{Probability Model Prediction}\label{sec:net_arch}

The proposed probability model prediction is designed to calculate the probability mass function (PMF) that statistically represents the occupancy symbols of the current coding unit, based on the 10 most recent PMFs previously computed. For that purpose, a three-layered neural network is employed. The proposed neural network architecture is relatively simple and has as input the 10 prior PMFs. The first layer is a custom fully connected FC($N$,$M$) layer, where $N$ is the number of input features in the input data, and $M$ is the number of channels (or neurons) in the output. In this layer, the input features correspond to all values associated to the same position in the PMF input vectors. Thus, this custom FC layer consists of 16 separate FC(10, 1) units, each followed by a ReLU activation function. The outputs of this custom FC layer are then passed into a fully connected FC(16, 64) layer, followed by another ReLU activation. The final fully connected FC(64, 16) layer produces the output, which is passed through a softmax activation function to ensure a valid probability distribution. This low-complexity architecture enables a quick generation of probability predictions, thus allowing the encoder and decoder to work within less than a \SI{1}{\micro\second} while improving the compression ratio compared to other non-learnable methods.

\subsection{Rice Coding}\label{sec:rice}

Rice coding is a parameterized Variable-Length Code (VLC) used in data compression, especially effective for encoding integer values. In this method, a parameter $k$ is chosen, and the data is divided into a quotient and remainder based on the chosen parameter. Therefore, for every source data element $n_i$ to encode, two values are computed with \eqref{eq:enc_rice1} and \eqref{eq:enc_rice2}
\begin{equation}\label{eq:enc_rice1}
q_i = \left\lfloor\frac{n_i}{2^k}\right\rfloor
\end{equation}
\begin{equation}\label{eq:enc_rice2}
m_i = n_i \bmod 2^k
\end{equation}
where $\left\lfloor p \right\rfloor$ denotes the floor function, which returns the greatest integer less than or equal to $p$, and mod stands for the modulo operation, which returns the remainder of a division. The encoding process of each source element $n_i$ is straightforward: first, $q_i$ is encoded in unary code, and then $m_i$ is encoded as an unsigned integer using $k$ bits. The unary code for $p$ is represented by $p$ bits set to 1, followed by a delimiter bit set to 0. The decoding follows the inverse process. The number of consecutive bits set to 1 is counted to obtain $q_i$. After discarding the delimiter bit, the subsequent $k$ bits represent $m_i$. The original value $n_i$ is then reconstructed using \eqref{eq:rec_rice}.
\begin{equation}\label{eq:rec_rice}
n_i = m_i + q_i \cdot 2^k
\end{equation}
This encoding method is particularly efficient for small non-negative integers and can be implemented using simple bit operations, eliminating the need for floating-point or integer calculations. These characteristics make Rice coding well-suited for applications that require fast and efficient compression of integer data.

\begin{figure}[h]
\includegraphics[scale=0.6]{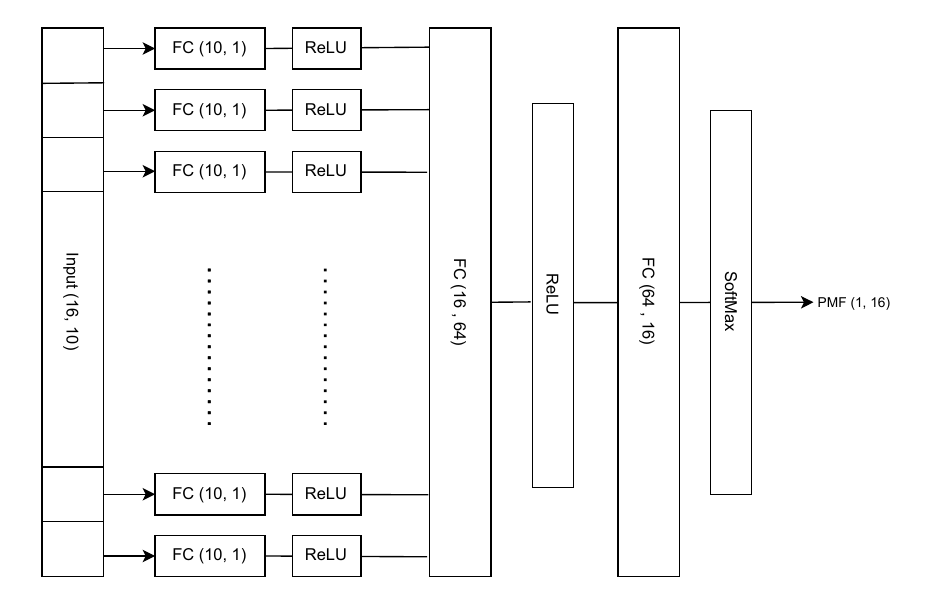}
\centering
\caption{Proposed neural network architecture for PMF prediction.}
\label{fig:network}
\end{figure}

% ===========================================================
% PERFORMANCE EVALUATION
% ===========================================================
\section{Performance Evaluation}\label{sec:performance}

\par This section presents the performance evaluation of the proposed LC-LLEC solution, covering the experimental results and their analysis, along with a concise description of the test material, performance metrics, and training procedure details.

\subsection{Test Material and Benchmarks}\label{sec:tmaterial}

\par The JPEG XE Common Test Conditions (CTC) \cite{JPEG-Data} establish a standardized framework for evaluating and comparing the performance of event-based codecs, ensuring a consistent and fair assessment across several solutions. Moreover, the JPEG XE reference dataset comprises a diverse collection of raw event sequences that capture a wide range of use cases, environmental conditions, durations, sensors, and sensor configurations. These sequences are encoded in the EVT2 format, with events organized chronologically according to their timestamps. In this work, nearly all JPEG XE sequences were selected for training, validation, and test, supplemented by additional sequences from \cite{psee-seq} to enhance the size of the test set. The selected sequences for training, validation, and test reflect two levels of sensor resolutions but more importantly, varying activity levels, quantified by the event rate (in million events per second - Mev/s); further details regarding the selected sequences are presented in Table~\ref{Table:allseq}.

Aligned with the JPEG XE CTC, the anchors selected are the lossless data codecs lz4 \cite{lz4}, bzip2 \cite{bzip2}, and 7z \cite{7zip}, which establish a good set of reference points for performance evaluation. All experiments were conducted with the full duration of the sequence, no temporal cropping was performed.

\begin{table}[ht]
    \scriptsize
    \centering
    \renewcommand{\arraystretch}{1.5} 
    \caption{Event sequences selected for training, validation and test.}
    \begin{tabular}{lccc}
        \hline
        Sequence name & Spatial Res. & Duration (s) & Event rate (Mev/s) \\
        \hline
        \multicolumn{4}{c}{\textsc{Training Sequences}} \\
        \hline
        Industrial\_spray & $640\times480$ & 1.6 & 1.30 \\
        Surveillance\_startracking & $1280\times720$ & 12.3 & 1.84 \\
        Deblur\_street & $1280\times720$ & 2.4 & 64.6 \\
        Traffic\_monitoring & $640\times480$ & 28 & 0.64 \\
        \hline
        \multicolumn{4}{c}{\textsc{Validation Sequences}} \\
        \hline
        Depthsensing\_highspeedlaser & $1280\times720$ & 3.0 & 38.8 \\
        Eyetracking\_right & $1280\times720$ & 9.9 & 5.75 \\
        \hline
        \multicolumn{4}{c}{\textsc{Test Sequences}} \\
        \hline
        Localization\_cube & $640\times480$ & 10.0 & 8.29 \\
        Activemarkers\_handheld & $1280\times720$ & 10.1 & 0.60 \\
        Industrial\_counting & $640\times480$ & 6.3 & 0.73 \\
        Industrial\_fluidflow & $1280\times720$ & 5.6 & 1.88 \\
        \textit{Spinner} & $640\times480$  &  5 &    10.83   \\
        \textit{Hand\_spinner} &   $640\times480$ &  5 &   2.39  \\
        \textit{195\_falling\_particles} & $640\times480$  & 0.032 &   632.52 \\
        \textit{Monitoring\_40\_50hz }&  $640\times480$ & 6  &  6.77  \\ 
        \textit{Cube}&  $640\times480$ & 17.7  &  7.00  \\ 
        \hline
    \label{Table:allseq}
    \end{tabular}
\end{table}

\subsection{Performance Metrics}

\par For performance evaluation, three objective metrics are used, as adopted by the JPEG XE CTC \cite{JPEG-Data}: the compression ratio (CR), the average compressed event size (S), and the runtime complexity (RC).

\par The CR metric quantifies the reduction in the size of the compressed bitstream generated by the proposed LC-LLEC solution (in bits) relative to the size (in bits) of the input sequence represented in the EVT2 format, as defined in \eqref{equ:cr}. The EVT2 format is a popular data representation for event sensors, where each event is typically encoded using 32 bits. It is a very robust format that does not contain any aggregation (contrary to EVT 3.0) and is destined for low-event rate applications.

\begin{equation}\label{equ:cr}
CR = \frac{\text{size}(\text{input sequence})}{\text{size}(\text{compressed bitstream})}
\end{equation}

\par The S metric is an additional measure of compression efficiency, quantifying the average number of bits the proposed LC-LLEC solution uses to represent a single event. It is calculated as the ratio of the compressed bitstream size (in bits) to the total number of events in the input sequence and is expressed in bits per event, as shown in \eqref{equ:s}.

\begin{equation} \label{equ:s}
S = \frac{\text{size}(\text{compressed bitstream})}{\text{number of input events}}
\end{equation}

\par Additionally, the runtime complexity metric RC assesses the computational efficiency of the proposed LC-LLEC solution excluding the quadtree creation, which can be done efficiently within the sensor. It measures the time taken to encode and decode the input sequence, separately, providing information about the proposed LC-LLEC solution computational complexity. This metric is also crucial for evaluating if the compression method can be applied in real-time applications.

\subsection{Training Procedure}

To train the probability model prediction neural network, the Adam optimizer \cite{kingma2015adam} is used with a learning rate of ($10^{-4}$); a scheduler with a decay rate of 0.1 is also applied every 5 epochs. Batch normalization is applied before each ReLU activation to standardize the inputs to each layer. This helps keeping the inputs more stable during training, making the learning process faster and more efficient. This technique also adds a slight regularization effect, potentially improving the model's generalization. To predict the distribution of the occupancy of the next coding unit, the neural network uses the 10 preceding PMFs. The training process has 100 epochs; however, to mitigate overfitting, an early-stopping mechanism is implemented with the patience of 10 epochs, which monitors validation loss. The neural network training is conducted on a machine equipped with an NVIDIA GeForce RTX 3090 GPU. The validation sequences are used to fine-tune hyperparameters of the neural network, including the number of layers, batch size, and learning rate. Regarding the loss function, it is used the cross-entropy, which measures the error between the predicted and actual probability distributions. Specifically, the cross-entropy between two probability distributions is defined as follows:

\[
H(p, q) = -\sum_{x} p(x) \log(q(x))
\]

where:
\begin{itemize}
    \item \( p(x) \) is the true probability distribution of the data. It represents the actual distribution from which the data is drawn. 
    \item \( q(x) \) is the predicted probability distribution, which is the predicted PMF produced by the neural network. 
\end{itemize}

Therefore, cross-entropy quantifies the difference between two probability distributions, the true distribution \( p \) and the predicted distribution \( q \). A low cross-entropy value indicates that the predicted distribution is close to the true distribution.

\subsection{Experimental Results and Analysis}

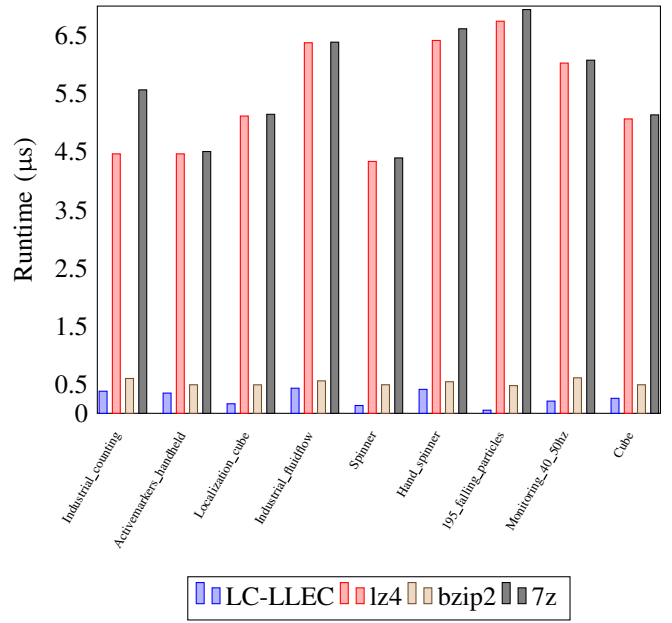
\begin{figure}[h]
\centering
\begin{tikzpicture}
\begin{axis}[
    ybar,
    bar width=3pt,
    width=0.5\textwidth,
    height=7cm,
    legend style={at={(0.5,-0.4)},
      anchor=north,legend columns=-1},
    ylabel={Runtime ($\upmu$s)},
    symbolic x coords={Industrial\_counting, Activemarkers\_handheld, Localization\_cube, Industrial\_fluidflow, Spinner, Hand\_spinner, 195\_falling\_particles, Monitoring\_40\_50hz, Cube},
    xtick=data,
    x tick label style={rotate=60,anchor=east,font=\tiny},
    ymin=0,
    ymax=7,
    ytick={0, 0.5,1.5,2.5,3.5,4.5,5.5,6.5},
    yticklabels={0, 0.5,1.5,2.5,3.5,4.5,5.5,6.5},
    minor y tick num=1,
    enlarge x limits=0.05,
    x tick style={draw=none},
    y tick style={draw=none},
    ]
\addplot coordinates {(Industrial\_counting,0.380) (Activemarkers\_handheld,0.348) (Localization\_cube,0.164) (Industrial\_fluidflow,0.432) (Spinner,0.134) (Hand\_spinner,0.412) (195\_falling\_particles,0.054) (Monitoring\_40\_50hz,0.210) (Cube,0.258)};
\addplot coordinates {(Industrial\_counting,4.46) (Activemarkers\_handheld,4.46) (Localization\_cube,5.11) (Industrial\_fluidflow,6.37) (Spinner,4.33) (Hand\_spinner,6.41) (195\_falling\_particles,6.74) (Monitoring\_40\_50hz,6.02) (Cube,5.06)};
\addplot coordinates {(Industrial\_counting,0.599) (Activemarkers\_handheld,0.491) (Localization\_cube,0.490) (Industrial\_fluidflow,0.558) (Spinner,0.491) (Hand\_spinner,0.543) (195\_falling\_particles,0.476) (Monitoring\_40\_50hz,0.610) (Cube,0.491)};
\addplot coordinates {(Industrial\_counting,5.56) (Activemarkers\_handheld,4.50) (Localization\_cube,5.14) (Industrial\_fluidflow,6.38) (Spinner,4.39) (Hand\_spinner,6.61) (195\_falling\_particles,6.94) (Monitoring\_40\_50hz,6.07) (Cube,5.13)};
\legend{LC-LLEC,lz4,bzip2,7z}
\end{axis}
\end{tikzpicture}
\caption{Encoding Runtime ($\upmu$s) per event.}
\label{fig:compression_runtime}
\end{figure}

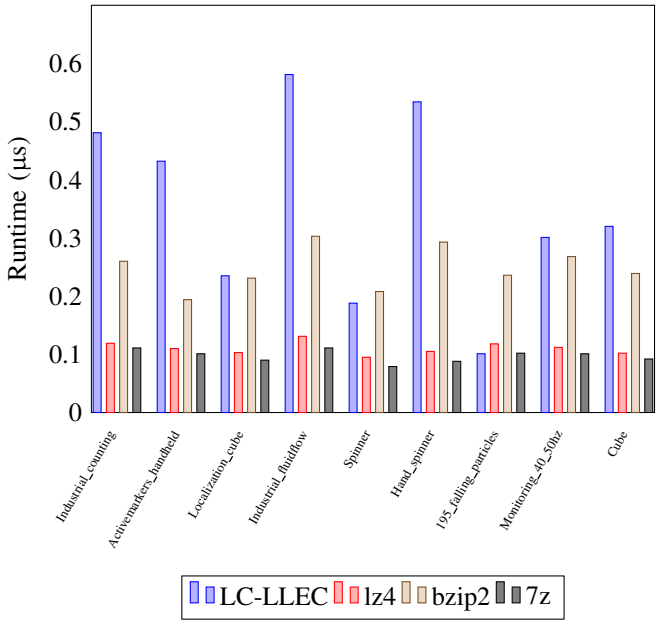
\begin{figure}[h]
\centering
\begin{tikzpicture}
\begin{axis}[
    ybar,
    bar width=3pt,
    width=0.5\textwidth,
    height=7cm,
    legend style={at={(0.5,-0.4)},
      anchor=north,legend columns=-1},
    ylabel={Runtime ($\upmu$s)},
    symbolic x coords={Industrial\_counting, Activemarkers\_handheld, Localization\_cube, Industrial\_fluidflow, Spinner, Hand\_spinner, 195\_falling\_particles, Monitoring\_40\_50hz, Cube},
    xtick=data,
    x tick label style={rotate=60,anchor=east,font=\tiny},
    ymin=0,
    ymax=0.7,
    ytick={0, 0.1, 0.2, 0.3, 0.4, 0.5, 0.6},
    yticklabels={0, 0.1, 0.2, 0.3, 0.4, 0.5, 0.6},
    enlarge x limits=0.05,
    x tick style={draw=none},
    y tick style={draw=none},
    ]
\addplot coordinates {(Industrial\_counting,0.481) (Activemarkers\_handheld,0.432) (Localization\_cube,0.235) (Industrial\_fluidflow,0.581) (Spinner,0.188) (Hand\_spinner,0.534) (195\_falling\_particles,0.101) (Monitoring\_40\_50hz,0.301) (Cube,0.320)};
\addplot coordinates {(Industrial\_counting,0.119) (Activemarkers\_handheld,0.110) (Localization\_cube,0.103) (Industrial\_fluidflow,0.131) (Spinner,0.095) (Hand\_spinner,0.105) (195\_falling\_particles,0.118) (Monitoring\_40\_50hz,0.112) (Cube,0.102)};
\addplot coordinates {(Industrial\_counting,0.260) (Activemarkers\_handheld,0.194) (Localization\_cube,0.231) (Industrial\_fluidflow,0.303) (Spinner,0.208) (Hand\_spinner,0.293) (195\_falling\_particles,0.236) (Monitoring\_40\_50hz,0.268) (Cube,0.239)};
\addplot coordinates {(Industrial\_counting,0.111) (Activemarkers\_handheld,0.101) (Localization\_cube,0.090) (Industrial\_fluidflow,0.111) (Spinner,0.079) (Hand\_spinner,0.088) (195\_falling\_particles,0.102) (Monitoring\_40\_50hz,0.101) (Cube,0.092)};
\legend{LC-LLEC,lz4,bzip2,7z}
\end{axis}
\end{tikzpicture}
\caption{Decoding Runtime ($\upmu$s) per event.}
\label{fig:decompression_runtime}
\end{figure}

Before presenting the overall experimental results obtained with the proposed solution, it is first analysed the selection of the Rice coding \textit{k} parameter. 
\begin{itemize}[leftmargin=*]

\item \textbf{Rice coding $k$ parameter selection}: To determine the best value of $k$, a comparative analysis of compression ratios across several sequences is performed. Table~\ref{tab:k_rice} presents compression ratios for different $k$ values (0, 1, 2, and 3) applied to a selected set of sequences. By examining the results, it is possible to identify the $k$ that consistently yields the highest compression ratio, which is clearly $k=1$ in this case (the value used for all other experiments). 

%============================================
% table for K of Rice coding
\begin{table}[h]
    \raggedleft
    \scriptsize
    \setlength{\tabcolsep}{3pt}
    \renewcommand{\arraystretch}{1.2}
    \caption{Compression ratios (CR) for rice coding $k$ selection}
    \begin{tabular}{|l|c:c:c:c|}
        \hline
        \textbf{Sequence name} & \textbf{CR ($k = 0$)} & \textbf{CR ($k = 1$)} & \textbf{CR ($k = 2$)} & \textbf{CR ($k = 3$)} \\
        \hline
        Localization cube & 1.60 & 2.29 & 2.02 & 1.63 \\
        Activemarkers handheld & 1.34 & 1.90 & 1.66 & 1.30 \\
        Industrial counting & 1.15 & 1.63 & 1.41 & 1.10 \\
        Industrial fluid flow & 0.90 & 1.25 & 1.07 & 0.83 \\
        Spinner & 1.98 & 2.88 & 2.67 & 2.21 \\
        Hand spinner & 1.05 & 1.48 & 1.26 & 0.98 \\
        195 falling particles & 1.63 & 2.40 & 2.26 & 1.85 \\
        Monitoring\_40\_50hz & 1.31 & 1.86 & 1.63 & 1.30 \\
        Cube & 1.60 & 2.28 & 2.02 & 1.63 \\
        \hline
    \end{tabular}
    \label{tab:k_rice}
\end{table}

Table \ref{table:CR_Metric} and Table \ref{table:S_Metric}, along with Figure \ref{fig:compression_runtime} and Figure \ref{fig:decompression_runtime} present the experimental results obtained with the proposed LC-LLEC solution compared to benchmarks. These results include compression ratio (CR), average compressed event size (S) metrics, runtime complexity (RC) for both the encoder and decoder, and the computational complexity of the probability model prediction, common to both encoder and decoder. The test sequences, detailed in Table \ref{Table:allseq} represent a wide range of scenarios (as discussed in Section \ref{sec:tmaterial}) and are essential for a comprehensive evaluation of LC-LLEC's performance. Based on the results obtained, the following conclusions can be drawn:
\item \textbf{Compression ratio (CR)}: The proposed LC-LLEC method demonstrates superior compression efficiency compared to the three benchmark algorithms, lz4, bzip2, and 7z, in terms of the CR metric, as shown in Table \ref{table:CR_Metric}. Across the nine test sequences, LC-LLEC consistently achieves higher compression ratio values, ranging from 1.25 to 2.88. In contrast, lz4 performs poorly with CR values predominantly below 1, while bzip2 shows moderate improvement (with CR between 1.10 and 1.51). The 7z algorithm proves to be the closest competitor, achieving CR values similar to LC-LLEC but still lagging in every test sequence, with a maximum CR reduction of 39.6\%. The LC-LLEC exhibits exceptional performance, particularly on the 'Spinner' sequence (CR of 2.88), significantly outperforming the next best result (7z with a CR of 2.59).

\begin{table}[h]
\scriptsize
  \centering
  \renewcommand{\arraystretch}{1.2}
  \caption{Compression Efficiency Results for the CR Metric.}
  \begin{tabular}{|c|*{4}{c|}}
    \hline
    \multirow{2}{*}{\bfseries Sequence name} & \multicolumn{4}{c|}{\bfseries CR} \\
    \cline{2-5}
     & \textbf{LC-LLEC} & \textbf{lz4} & \textbf{bzip2} & \textbf{7z} \\
    \hline
    Industrial\_counting & 1.63 & 0.93 & 1.28 & 1.56 \\
    \hline
    Activemarkers\_handheld & 1.90 & 0.90 & 1.27 & 1.86 \\
    \hline
    Localization\_cube & 2.29 & 0.99 & 1.25 & 1.84 \\
    \hline
    Industrial\_fluidflow & 1.25 & 0.94 & 1.10 & 1.14 \\
    \hline
    Spinner & 2.88 & 0.99 & 1.51 & 2.59 \\
    \hline
    Hand\_spinner & 1.48  & 0.97 & 1.35 & 1.45 \\
    \hline
    195\_falling\_particles & 2.40 & 0.99 & 1.21 & 1.45\\
    \hline
    Monitoring\_40\_50hz & 1.86 & 0.99 & 1.26 & 1.54 \\ 
    \hline
    Cube & 2.28 & 0.99 & 1.25 & 1.84 \\ 
    \hline
  \end{tabular}
  \label{table:CR_Metric}
\end{table}

\item \textbf{Average compressed event size in bits (S)}: LC-LLEC consistently outperforms lz4, bzip2, and 7z compression algorithms across all test sequences in terms of average compressed event size (S metric), as shown in Table~\ref{table:S_Metric}. LC-LLEC achieves the lowest S values for each sequence, indicating superior compression efficiency. The improvement is particularly notable compared to lz4 and bzip2, with LC-LLEC often reducing the compressed size by 40-50\% relative to these methods. Even when compared to the generally more competitive 7z algorithm, LC-LLEC still manages to achieve lower S values in all cases, showcasing its effectiveness in minimizing the average compressed event size for event sequences with rather different characteristics.

\begin{table}[h]
  \scriptsize
  \centering
  \renewcommand{\arraystretch}{1.2}
  \caption{Compression Efficiency Results for the S Metric.}
  \begin{tabular}{|c|*{4}{c|}}
    \hline
    \multirow{2}{*}{\bfseries Sequence name} & \multicolumn{4}{c|}{\bfseries S} \\
    \cline{2-5}
     & \textbf{LC-LLEC} & \textbf{lz4} & \textbf{bzip2} & \textbf{7z} \\
    \hline
    Industrial\_counting & 19.62  & 34.26   & 24.93  & 20.41 \\
    \hline
    Activemarkers\_handheld & 16.79  & 35.17  & 25.10  & 17.11 \\
    \hline
    Localization\_cube & 13.96 & 32.24   & 25.51 & 17.31 \\
    \hline
    Industrial\_fluidflow & 25.41 & 33.71 & 29.05 & 27.90 \\
    \hline
    Spinner & 11.09 & 32.17 & 21.14 & 12.32 \\
    \hline
    Hand\_spinner & 21.61  & 32.83 & 23.58 & 22.00 \\
    \hline
    195\_falling\_particles & 13.32 & 32.00 & 26.31 & 21.95 \\
    \hline
    Monitoring\_40\_50hz & 17.15 & 32.29 & 25.25 & 20.72 \\ 
    \hline
    Cube & 13.98 & 32.28 & 25.58 & 17.34 \\ 
    \hline
  \end{tabular}
  \label{table:S_Metric}
\end{table}

\item \textbf{Runtime comparison per event (RC)}: The proposed LC-LLEC method achieves lower encoding runtime across all test sequences when compared to lz4, bzip2, and 7z, as shown in Figure \ref{fig:compression_runtime}. More precisely, LC-LLEC consistently achieves the fastest encoding times, with encoding runtimes ranging from \SI{0.054}{\micro\second} to \SI{0.432}{\micro\second}  per event. This is faster than the next best solution, bzip2, whose encoding runtime ranges from \SI{0.476} to \SI{0.610} {\micro\second} per event. Both lz4 and 7z show notably slower encoding times, often exceeding \SI{4}{\micro\second}  per event. However, in terms of decoding, the LC-LLEC method is generally slower than the other three algorithms, as Figure \ref{fig:decompression_runtime} shows. While LC-LLEC's decoding times range from \SI{0.101}{\micro\second} to \SI{0.581}{\micro\second} per event, the other methods, particularly lz4 and 7z, achieve faster decoding speeds, often below \SI{0.120}{\micro\second} per event. This trade-off indicates that LC-LLEC achieves fast encoding at the expense of slower decoding, which can be beneficial in situations where quick data encoding is essential, and decoding speed is less of a priority. For instance, one of the most promising use cases for event data today involves connecting resource-constrained sensor devices to application platforms with greater resources. Additionally, it’s important to highlight that formats like lz4, bzip2, and 7z are designed for data storage rather than streaming, allowing them to exploit correlations across the entire sequence, something that is not feasible in real-time streaming applications.

\item \textbf{Computational complexity per event}: The computational complexity of the probability model is shown in Table \ref{table:KMacTable}, which presents the Kilo Multiply-Accumulate Operations per Cycle (kMAC) operations per million of events for several sequence types. The kMAC/MEvents ranges from 0.039 to 1.056, indicating extremely low computational requirements across different scenarios. For instance, the 'Industrial\_counting' sequence, which has the highest complexity, only requires 1.056 kMAC/MEvent, while the 'Cube' sequence only requires 0.039 kMAC/MEvent. This low computational complexity makes the model particularly suitable for applications where resources are limited or where real-time processing is crucial. The consistently low kMAC/MEvent values across diverse sequence types also suggest that the model maintains its efficiency regardless of the input data characteristics as expected, further highlighting its versatility and practical applicability in various scenarios.

\end{itemize}

% =================================================================
% Runtime

\begin{center}
\begin{table}[H]
  \centering
  \renewcommand{\arraystretch}{1.2}
  \caption{Probability model prediction computational complexity (KMac/MEvent).}
  \begin{tabular}{|c|*{1}{c|}}
    \hline
    \textbf{Sequence name} & \textbf{Complexity (KMac/MEvent)} \\
    \hline
    Industrial\_counting & 1.056   \\
    \hline
    Activemarkers\_handheld & 0.798  \\
    \hline
    Localization\_cube & 0.058  \\
    \hline
    Industrial\_fluidflow & 0.458  \\
    \hline
    Spinner & 0.089  \\
    \hline
    Hand\_spinner & 0.405   \\
    \hline
    195\_falling\_particles & 0.239 \\
    \hline
    Monitoring\_40\_50hz & 0.119  \\ 
    \hline
    Cube & 0.039  \\ 
    \hline
  \end{tabular}
  \label{table:KMacTable}
\end{table}
\end{center}

\section{Conclusions and Future Work}\label{sec:conclusions}
This work presents a new low-complexity, learning-based lossless event data compression framework, exploiting a quadtree data structure to represent events. Moreover, the proposed probability model prediction based on a neural network, combined with a frequency substitution table, boosts the effectiveness and efficiency of the Rice entropy coding solution. Experimental results demonstrate that the proposed approach consistently outperforms existing lossless compression techniques across sequences with varying activity levels. Additionally, the proposed method presents an impressive computational efficiency and runtime performance, highlighting its practical applicability. Other advantages of the proposed solution is the capability to easily provide random access points via bit stuffing. 

As future work there are several areas that may further bring coding efficiency benefits without compromising encoder complexity. The prediction probability model may be improved with quantized neural networks that use integer parameters and activations (weights and bias), for example, W8A16 where 8 bits are used for weights and 16 bits for activations, this will lower the computational power demands of this part of the encoder. Another possibility is to perform entropy coding of the polarity vector which is now sent uncompressed or perform a more fine adjustment of $k$, e.g. for each coding unit.

\bibliographystyle{ieeetr}
\bibliography{Ref}
\end{document}